\documentclass[%
reprint,
amsmath,amssymb,
pre,
]{revtex4-1}
\usepackage{bbold}
\usepackage{graphicx}
\usepackage{bm}
\usepackage{hyperref}
\usepackage[mathlines]{lineno}

\begin{document}
\title{Hypergraphs and City Street Networks}
\author{Thomas Courtat$^{1,2}$, Catherine Gloaguen$^1$, Stephane Douady$^2$}
\email{thomas.courtat , catherine.gloaguen (at) orange-ftgroup.com; douady (at) lps.ens.fr}
\affiliation{$^1$Orange Labs, 38-40, rue du G\'en\'eral Leclerc, 92794 Issy-les-Moulineaux, France \\ $^2$Laboratoire Mati\`ere et Syst\`emes Complexes (MSC), UMR CNRS
-Universit\'e Paris Diderot CC 7056, 10 rue Alice Domon et L\'eonie Duquet, 75205 Paris Cedex 13}
\date{\today}
\begin{abstract}
The map of a city's streets constitutes a particular case of spatial complex network. However a city is not limited to its topology: it is above all a geometrical object whose particularity is to organize into short and long axes called streets. In this article we present and discuss two algorithms aiming at recovering the notion of street from a graph representation of a city. Then we show that the length of the so-called streets scales logarithmically. This phenomenon leads to assume that a city is shaped into a logic of extension and division of space. 
\end{abstract}
\maketitle
\section{Introduction}
Traditionally the map or equivalently the street network of a city is represented by a graph with portion of streets considered as edges and their intersections as vertices. Since such a graph is large (7000 vertices for an average city) and displays non trivial patterns it came to the complex systems \cite{Albeverio2008} and complex networks \cite{Boccalettia2006,Barthelemy2010} field of study. In \cite{Blanchard2009} the topology of this graph is studied by means of random walks, \cite{Jiang2004,Cardillo2006,Porta2006,Buhl2006} study classical complex network parameters and \cite{Crucitti2006} introduces spatiality to its work by means of shortest path distances and the notion of \textit{centrality}. 
\\
However this purely topological representation does not take into consideration the whole geometrical information of a city. In this article we define geometrical and straight graphs plus an integral allowing handling with a city as a geometrical object, the graph structure being only a skeleton that holds it up. The geometry of street segments is yet particular. They are coherently arranged into disjoint geometrical sets: the streets. We seek out from plain vector maps (i.e. vector collections of street segments) to recover the notion of street and thus to get a multi-scale representation of the city. At this point, one can really speak of street networks, we mathematically represent by \textit{straight hypergraphs}.
\\
The street appears as a turn in the notion of axes and visibility graph used in the Space Syntax framework \cite{Hillier1984,Hillier1993,Hillier2002,Albeverio2008}. The visibility map is not robustly defined with respect of small variations on a map. It is very sensitive to local curvature and to the sampling of the map \cite{Ratti2004}. Various method have been proposed to overcome this inconsistency. The notion of axes is replaced in \cite{Jiang2002} by the notion of named-street: two axes are the same if they have the same name in the data basis. In \cite{Figueiredo2005} two axes are melt if their angle is less or equal than a threshold ($45^{\circ}$). But the resulting set of streets depends on the starting point of the algorithm. The Intersection Continuity Principle is presented in \cite{Porta2006}: two axes are melt at an intersection if they make the largest convex angle between all angles at the intersection. 
\\
The originality of our approach is to define streets formally in a framework devoted to cities, propose two algorithms computationally optimized and check their agreement with reality.
\\
In a first part we introduce a formal framework to represent cities as both topological and geometrical objects. Then we present two algorithms depending on a single parameter to partition \textit{street segments} into \textit{streets}. From a data basis of 109 (not truncated) French towns we tune this parameter and asses the performances of each algorithm. To end with we study the resulting distribution of street lengths. We statistically prove from our data basis that street lengths in a city follow a mixture of log-normal laws and interpret this as the result of an extension / division of space process. 
 
\section{Formal representation of city maps}
We represent a city by the notion of geometrical and straight graph. The vocabulary in use is freely adapted from general graph and geometric graph theory \cite{Gross2004}. The notion of straight graph directly corresponds to the one of planar straight line graph. The main difference is the point of view we adopt and the topological and differential structures we provide on the set of geometrical graph, see \cite{Courtat2011,Courtat2011a} for details. 
\subsection{Geometrical graphs}
A graph $G = (V,E)$ is a finite number of vertices $V$ and a part $E$ of $V \times V$. 
If $\sharp V$ is large one would prefer to use the word network.
If $V$ are points in an Euclidian space we speak of spatial networks \cite{Barthelemy2010} and if elements of $E$ are materialized by geometrical curves that intersect only at their extremities that are elements of $V$ we will say here that we have a \textit{geometrical graph}. 
Hence a geometrical graph is both a topological object ( from $(V,E)$) and a geometrical one (elements of $E$ are curves). When elements of $E$ are segments, we will say $G$ is a \textit{straight graph}. $V= (v_1, ... ,v_n) \in (\mathbb{R}^{2})^n$ and it is totally definded by its adjacency matrix $A =(a_{ij})$. 

\subsection{Hypergraph additional structure}
A \textit{hypergraph} is a graph whose edges can contain more than two nodes. 
If $G = (V,E)$ is a graph and $R$ an equivalence relationship on $E$ then the set of equivalence classes $E/R$ constitutes hyper-edges: $(V, E/R)$ is a hypergraph.
In a urban context we can think of $R=$ "these edges have the same street name". We present below two relationships that define the city hypergraph structure $H$ directly from the spatial information of $G$ without additional data. We write $G = ((V,E),H)$
\subsection{City graphs}
A city graph is a straight graph representation the street network of a city. This kind of graph has particular features studied for instance in \cite{Buhl2006}. 
A city graph writes $C= ((V,E),H)$ where $(V,E)$ is a straight graph and $H$ an additional hypergraph structure. 
Elements of $E$ are called street segments, they have no physical meaning: they are a sampling of the network. Elements of $H$ are called streets.
\\
The degree is a function defined on $V$ that associates to each vertex the number of edges that pass trough it.
We write $V = V_1 \cup V_2 \cup V_+$ with $V_1$ vertices of degree 1 called dead-ends, $V_2$ vertices of degree 2 called junctions (and seen as sampling artifacts) and vertices of degree $\geqslant 3$ intersections. 
$d°$ can be extended to each point on an edge: $\forall e \in E, \forall x \in \mathrm{Int}(e), d°_C(x) = 2$. $(V,E)$ is a particular skeleton of $C$, any point in the interior of an edge can be added as an element of $V_2$ without changing the overall structure.
If $e \in E$, $V(e)$ is the set of extremities of $e$ in $V$. If $v \in V$, $E(v)$ is conversely the set of edges passing through $v$ and if $v \in V(e)$, $v(e)$ is the other extremity of $e$. 
An element $h\in H$ can be seen as a subgraph of $C$ and induces a degree function $d°_h$.
\subsection{Data}
Maps are imported from a data basis of French regions vector maps "ESRI". A set of 109 cities is extracted. For each of them we get a geometry file ".MIF" and an attribute table ".mdb".
\\
The geometry of the street system is coded by a list of poly-lines. We underscan ".MIF" by taking care of preserving the angles at the intersections. We create a structure $V$ containing the position of each vertex and a structure $E$ containing for each edge two references to $V$ for its extremities. $H$ is an array with as many element as there are in $E$. Each element is a "label" (an integer) coding for the hyperedge to which belongs the edge.
\\
The result is noisy with detached structures (about $5$ percent) we erase by only keeping the largest connected component of the graph. We also erase edges appearing several times. For some algorithms its is more efficient to change the representation of the graph. For instance we can change $E$ to an adjacency matrix or and adjacency lists (a list for each vertex of the edges passing through it and another list of adjacent vertices).
\\
The attribute table focuses on street segments with additional information such as length and name (the same name is attributed to street segments that compose the same "named-street"). We will see this table is more indicative than trustable. 

\section{Two algorithms to recover streets}
Let $C=(V,E)$ be a city graph. To compute a $H$ structure we will use the following property: 
If $\hat{R}$ is a reflexive relationship on $E^2$ then the relationship $R$ on $E$ defined by: 
\begin{equation} 
e_1 \, R \, e_2 \ \mathrm{iif} \ \exists \: \alpha_1=e_1, \, \alpha_2,\, ... \, , \, \alpha_n = e_2 \in E \, | \\ \alpha_1 \ \hat{R} \ \alpha_2, \, \alpha_2 \ \hat{R} \ \alpha_3, .... , \alpha_{n-1} \ \hat{R} \ \alpha_n \label{chaine} \end{equation} 
is an equivalence relationship (transitive closure).
\\
Notice that defying a Hypergraph via an equivalence relationship provides an algorithm not depending on its starting point.
\subsection{Angular tolerance (AT)}
\begin{figure*}
\begin{center}
\includegraphics[scale = 0.3 ]{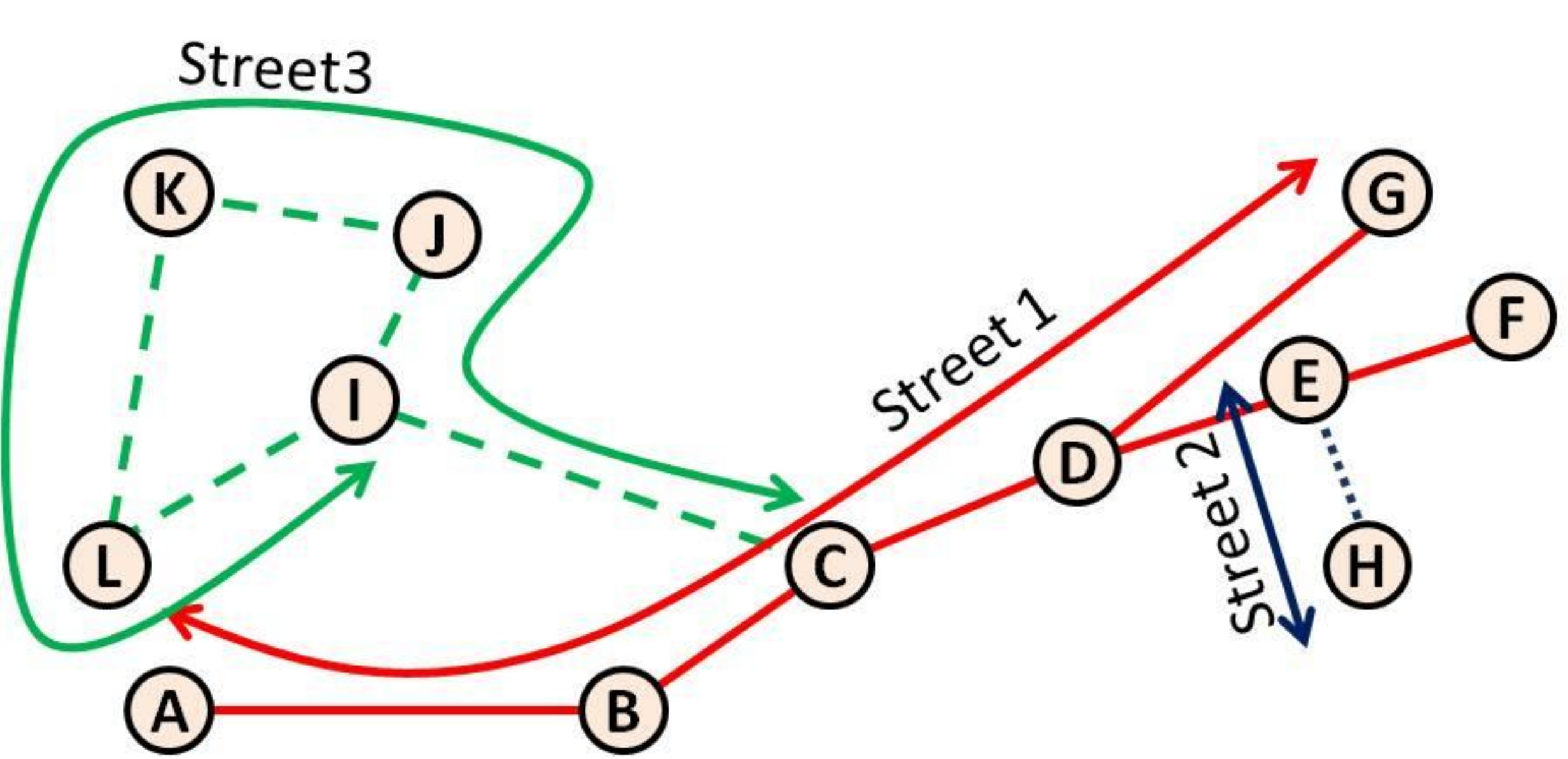}
\includegraphics[scale = 0.5 ]{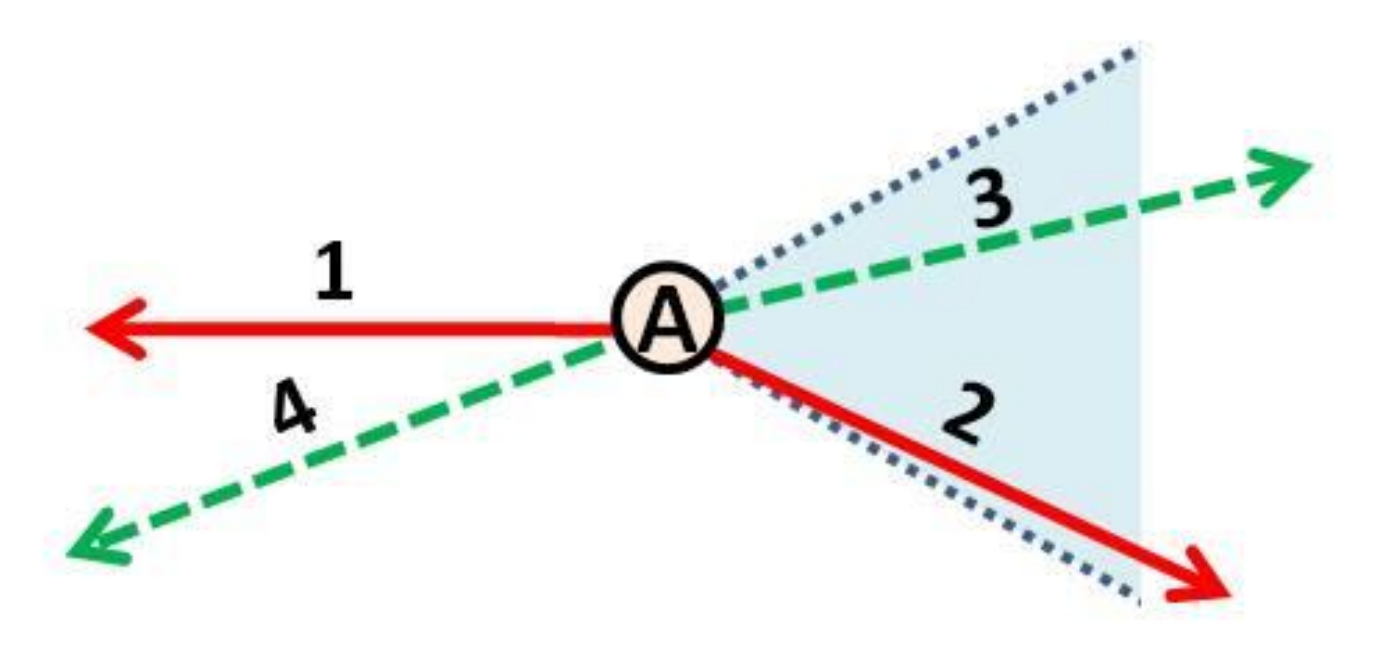}
\caption{\label{hypergraph} LEFT: Street 1 is branched (vertex $D$) and the Street 2 contains a loop. RIGHT: At intersection $A$, segment $1$ could be associated to $3$ and $2$. The closest angle to $\pi$ is made by $3$ but $3$ and $4$ correspond to an angle reciprocally minimal. They are associated and $1$ goes with $2$. The same reasoning leads to the same associations whatever the first segment considered.}
\end{center}
\end{figure*}

\begin{figure}
\begin{center}
\includegraphics[scale = 0.4]{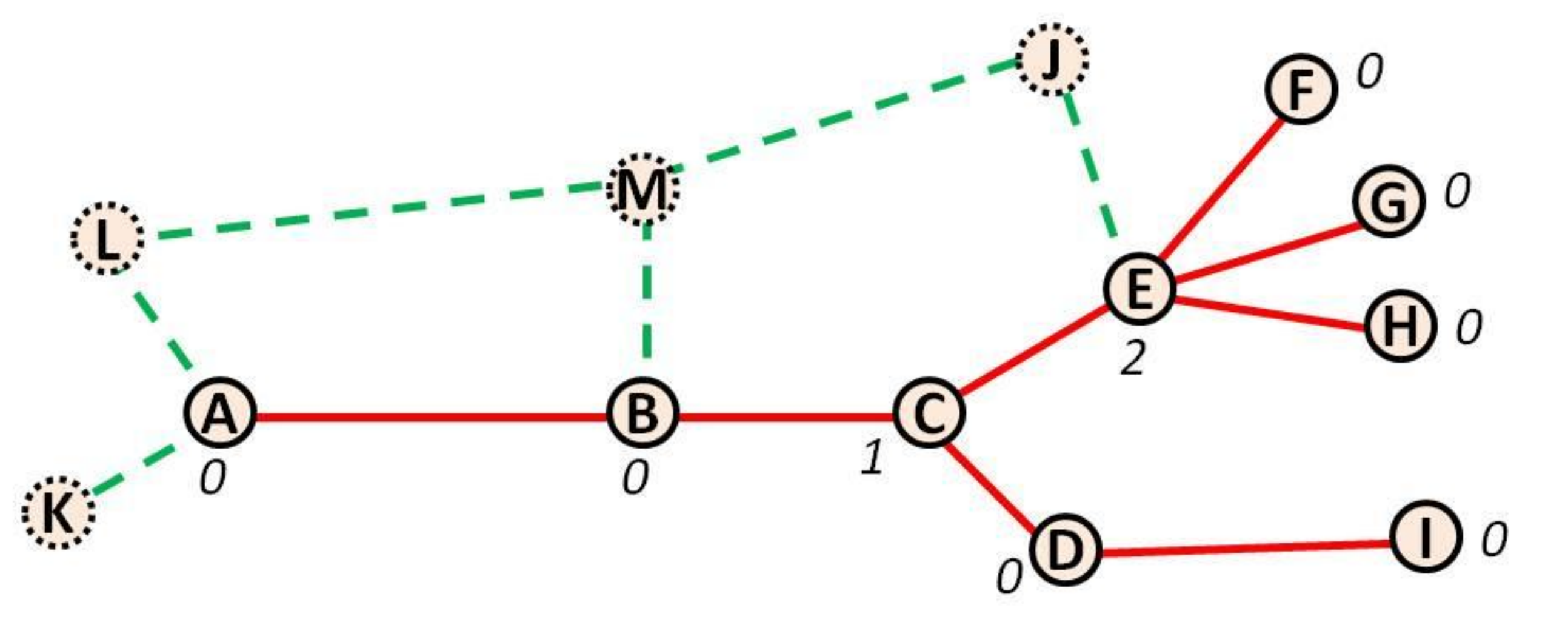}
\caption{\label{branches} The red solid graph is a street. Its number of branches is $1+2=3$.}
\end{center}
\end{figure}

We use the reflexive relationship $\hat{R}_{\theta}$ depending on the angular parameter $\theta$: 
\begin{multline}
e_1 \ \hat{R}_\theta \ e_2 \quad \mathrm{iif}\quad \exists v, v_1, v_2 \in V, e_1=[v v_1] , e_2 = [v v_2], \quad \\ (d(v) = 2) \vee ( |(\measuredangle(\overrightarrow{v v_1},\overrightarrow{v v_2})-\pi | \leq \theta)) 
\end{multline}
this relation considers that two adjacent street segments are part of the same street if they meet at a junction or if they meet at an intersection but remain almost aligned (Fig. \ref{hypergraph} left). 
This algorithm strongly risks producing "branched streets" (red solid line in Fig. \ref{hypergraph} left, Fig. \ref{branches}).
\subsection{The minimal reciprocal alignment (MRA)}
To define $\hat{S}_{\theta}$ we position at particular vertex $v$ and consider the set of the edges passing through it $E(v)=\{ e_1 =[v_1v],.., e_n=[v_nv] \}$. 
We iteratively define $\hat{S}_{\theta}$ with the variable $s$: 
(1) the initial "remaining edges" is set for $s=0$: $E_0 = E(v)$,
(2) we consider all pairs of edges $(e_i ,e_j) \quad 1 \leq i<j \leq n $, $ e_i S_{\theta} e_ j$ iif 
\begin{multline}
|(\measuredangle(\overrightarrow{v v_j},\overrightarrow{v v_j})-\pi | \leq \theta) \\ \text{ and }  \forall e_k=[v v_k] \in E_s \neq e_i,e_j \text{ , } |(\measuredangle(\overrightarrow{v v_k},\overrightarrow{v v_i})-\pi | < |(\measuredangle(\overrightarrow{v v_i},\overrightarrow{v v_j})-\pi | \\ \text{ and } |(\measuredangle(\overrightarrow{v v_k},\overrightarrow{v v_j})-\pi | < |(\measuredangle(\overrightarrow{v v_i},\overrightarrow{v v_j})-\pi |
\end{multline}
 Two edges are associated if they are the most aligned in $E_s$.
(3) $E_{s+1}$ is $E_s$ without the edges associated in the $s$ step. 
(4) We go on till $(E_s)$ stabilizes.

The reflexivity on the minimal condition induces the reflexivity of $\hat{S}_{\theta}$.
\\
For instance in Fig. \ref{hypergraph} right: $E_0 =\{1 ,2, 3, 4 \}$, $3$ and $4$ are associated, $E_1 = \{ 1, 2 \}$, $1$ and $2$ are associated and the algorithm ends. 

\subsection{Implementation}

Both algorithms can be implemented within the same skeleton by encapsulating two functions "Relation" with a boolean output, taking as parameters a vertex $v$, $E(v)$ and two distinct elements of it. The algorithm divides in two steps: (1) determine local relations between segments (2) transform this relation into equivalence classes by using Eq. \ref{chaine}. In the following code we mix up objects and their indice in an array.

\medskip
\noindent
\begin{small}
\begin{verbatim}
FUNCTION H = Hypergraph(Graph) 
    V = Graph.Vertices (v by 2 array)
    E = Graph.Edges (e by 2 array)
    H= new Array(e by 1)
    Cor = new Array(e by 10)
% {STEP 1 }
    FOR i= 1 to v
        EExtract = find e in E such i in e (E(i))
        FOR j < k 
            e1 = EExtract(j)
            e2 = EExtract(k)
            IF Relation(i, e1, e2, EExtract)
                Cor(e1, next available) = e2
                Cor(e2, next available) = e1
            END IF 
        END FOR 
    END FOR 
% {STEP 2}
    CurrentMark = 1
    FOR i = 1 to e
        IF H(i) = 0
            stak = [i]
            WHILE notEmpty(stak)
                current = pop(stak)
                H(current) = CurrentMark
                push(stak, set Cor(e , not = 0))
            END
            CurrentMark ++
            END IF
    END FOR
END FUNCTION
\end{verbatim}\end{small}
\noindent
With plain graph structure, the complexity is $O(v \times e)$ (Step 1) and $O(e)$ (Step 2) thus globaly in $O(v^2)$ (usualy $e \simeq 1.5v$). With an adjacency list (calculated in $O(e)$) Step 1 becomes $0(v)$ and the whole algorithm is $O(v)$.
\section{Tuning and Performances }
We have specified $AT$ and $MRA$ with a single angular parameter $\alpha$. In practice we want the algorithm to recover the actual streets of a city. 
It is hard to access to these information with our data: there are as many streets as there are different \textit{street names} in the data basis But in a particular city, their number can be extracted although not trustable. We just try to reach the true number of streets.
Add to that (AT) and to a lesser extent (MRA) risk to produce branched rather straight streets. We define \textit{the branching coefficient} to describe this tendency and seek out to minimize it.
\\
In this section we assess the performance of the algorithm and deduce an optimal tunning for $\alpha$ from a corpus of $N=109$ major French towns: $(C_1,..,C_N)$.
\subsection{Criteria}
\subsubsection{Number of street recovering}
We assume we know for $N$ cities their actual number of streets: $T_1, ..., T_N$.
Let $\alpha \longrightarrow f_k(\alpha)$ the function that associates to an angle $\alpha$ the number of streets one of our algorithm asses for the city $k$. If the algorithm is relevant, the quadratic error 
\begin{equation}
\Delta^2(\alpha) =\frac{1}{N}\sum_{k=1}^{N} \left( \frac{f_k(\alpha)-T_k}{T_k}\right) ^2 
\end{equation}
is small. However $(T_k)$ is not accurate. Some street segments have a blank "NAME" field. The data basis underestimates the number of streets. To get around this problem, we assume the error in the data basis is proportional to the proposed number of streets: $\tilde{T}_k = (1+ \lambda)T_k \quad \forall k$. The criterion rewrites in function of $\alpha$ and $\lambda$: 
\begin{equation}
\Delta^2(\alpha, \lambda) = \frac{1}{N}\sum_{k=1}^{N} \left( \frac{f_k(\alpha)-(1+\lambda)T_k}{(1+\lambda)T_k}\right)^2 
\end{equation}
A quick study of the data basis behavior permits to assess that $0.1<\lambda < 0.7$. 
$\frac{\partial C^2}{ \partial \lambda} (\alpha, \lambda) = 0 $ leads to a functional relationship between $\alpha$ and $\lambda= \lambda_{\alpha}$: 
\begin{equation}
\lambda(\alpha) = \frac{\sum f_i(\alpha)^2/T_i^2}{\sum f_i(\alpha)/T_i^2} -1
\end{equation}
and the criterion rewrites only in function of $\alpha$:
\begin{equation}
\Gamma^2(\alpha) =\frac{1}{N}\sum_{k=1}^{N} \left(1-\frac{f_k(\alpha)}{T_k.\frac{\sum f_i(\alpha)^2/T_i^2}{\sum f_i(\alpha)/T_i}}\right)^2 
\end{equation}

\subsubsection{Branching coefficient}
Let $H$ a hypergraph structure computed from $C$ and $h \in H$ a street, seen as an extracted subgraph of $C$. The number of branches in $h$ is defined by:
\begin{equation}
\xi(h) = \sum_{v \in h} \max (d°_h(i)-2, 0)
\end{equation} 
To measure the branched aspect of $H$ we define its branching coefficient from the number of branches of its streets:
\begin{equation}
\Xi(H)= \frac{\sum_{h \in H} \xi(h)}{\sum_{k>2}(k-2).d°(k)}
\end{equation}
If none of the streets is branched, $\Xi = 0$ and if $H$ is componed of a single street, $H$ is maximally branched wit $\Xi =1$.
\subsection{Analysis}
\subsubsection{AT} 
\begin{figure*}
\begin{center}
\includegraphics[scale = 0.4]{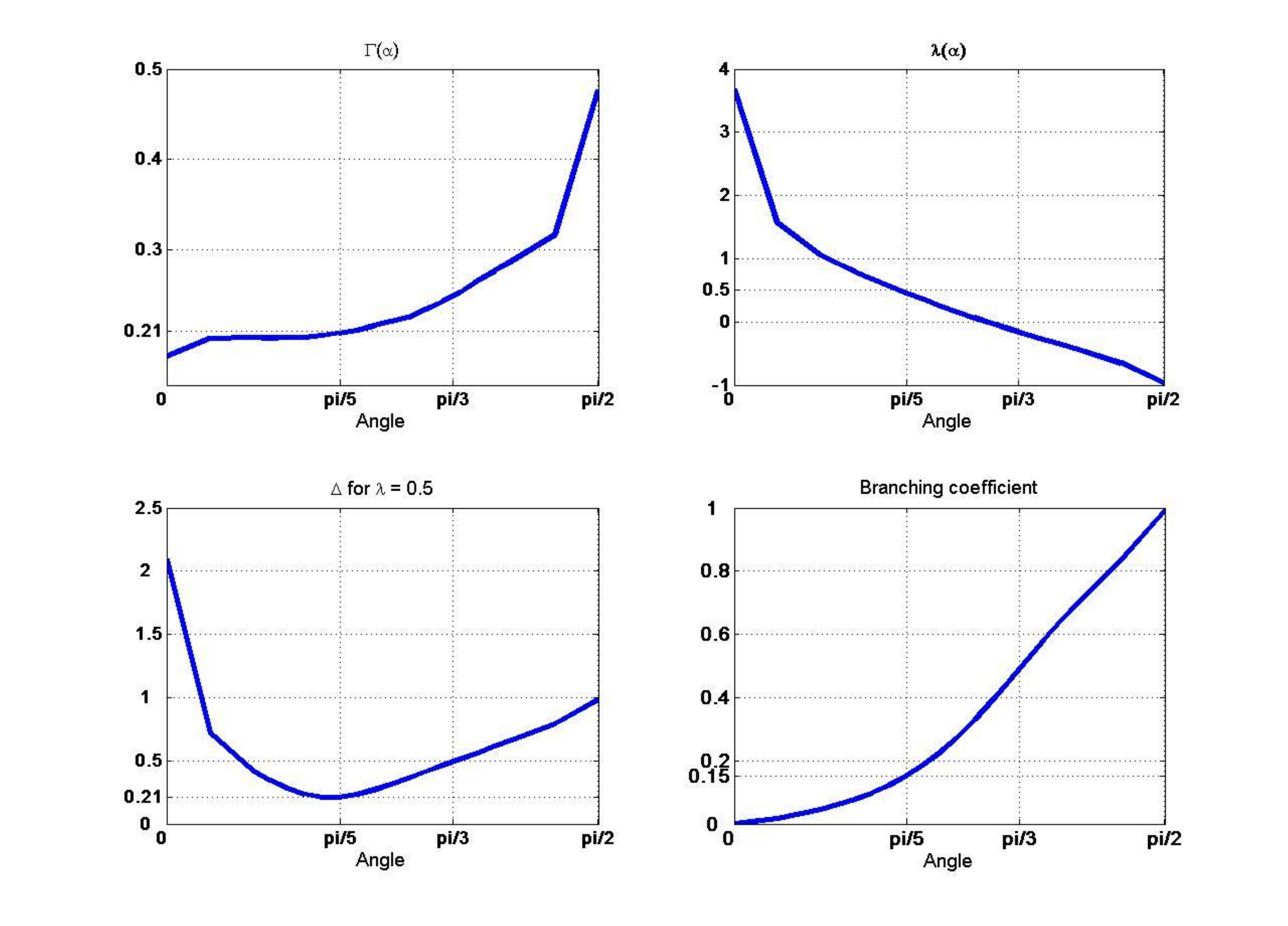}
\caption{\label{firstAlgo} Tuning and performances of the Angular Tolerance algorithm (AT). Top-left: $\Gamma(\alpha)$, Top-Right: $\lambda(\alpha)$, Bottom-Left: $\Delta(\alpha)$ for $\lambda=0.5 $ and Bottom-right: the increasing of the mean branching coefficient $\Xi$ with $\alpha$. }
\end{center}
\end{figure*}

The function $\Gamma(\alpha)$ reaches its minimum ($21 \% $) around $\pi/5$ (Fig. \ref{firstAlgo} top-left). This corresponds to a $\lambda = 0.5$ (Fig. \ref{firstAlgo} top-right) which is coherent with the order of weight we expressed. The abslute minimum in $0$ is eliminated since it drives to an aberrant value of $\lambda$. The branching coefficient is in average $\Xi=0.15$ which is slightly high but stays reasonable (Fig. \ref{firstAlgo} bottom-right). Fig. \ref{firstAlgo} bottom-left shows the criterion for $\lambda $ constant equal to $0.5$. With the corrected number of streets the criterion is convex and $\pi/5$ appears as a rather good and stable minimum.
\subsubsection{MRA} 
\begin{figure*}
\begin{center}
\includegraphics[scale = 0.4]{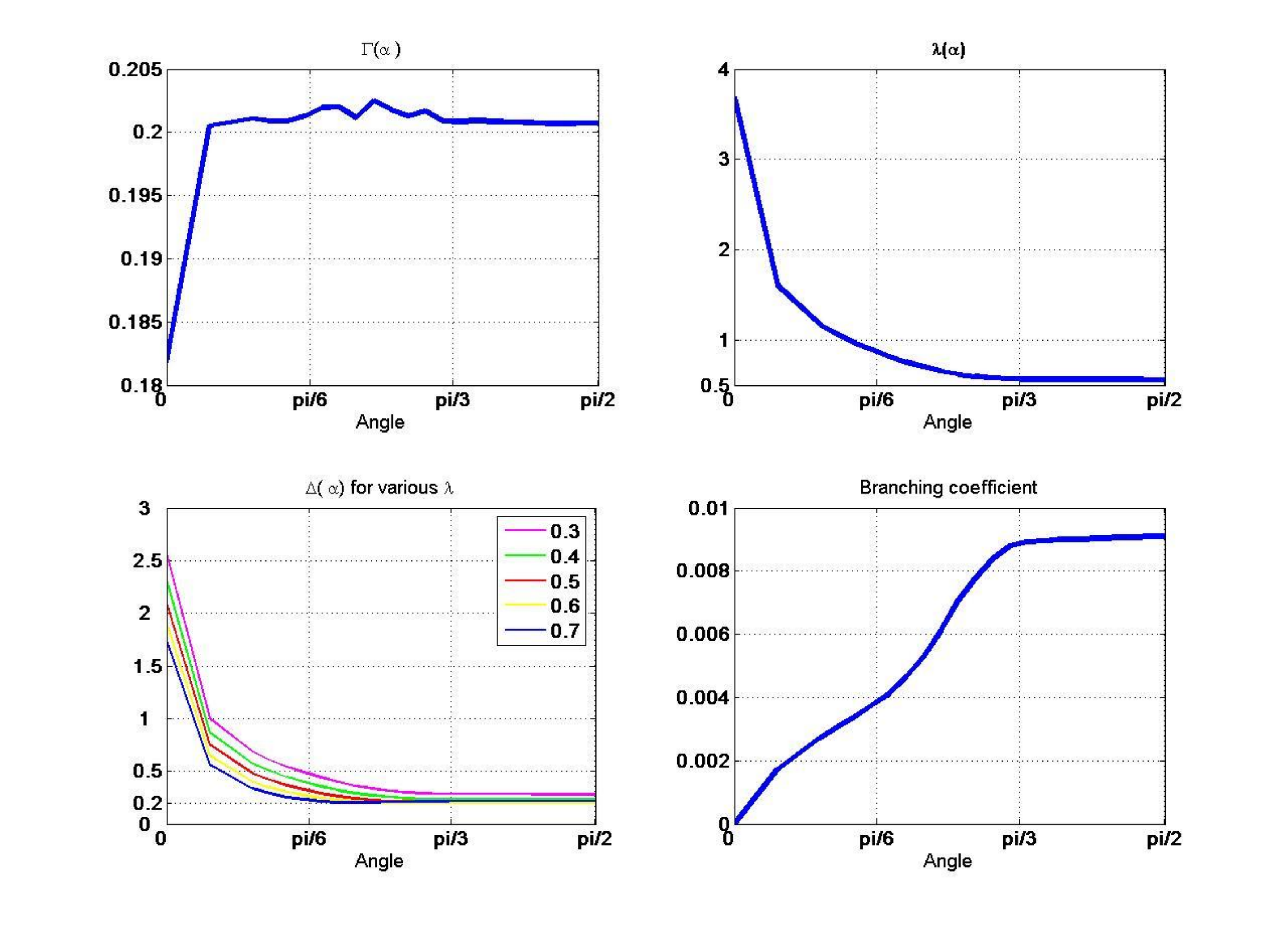}
\caption{\label{2Algo} Tuning and performances of the Minimal Reciprocal Angle algorithm (MRA). Top-left: $\Gamma(\alpha)$, Top-Right: $\lambda(\alpha)$, Bottom-Left: $\Delta(\alpha, \lambda)$ for $\lambda = 0.3$ to $0.7$ (its optimal value) and Bottom-right: the increasing of the mean branching coefficient with $\alpha$ asymptotically inferior to $1\%$ }
\end{center}
\end{figure*}

The function $\Gamma(\alpha)$ is almost constant equal to $0.2$ (Fig. \ref{2Algo} top-left). $\lambda_{\alpha}$ is exponentially decreasing with an asymptotic value of $0.56$ (Fig. \ref{2Algo} top-right). Added to that, $\forall \lambda \in [0, 1]$ $C(\alpha, \lambda)$ has an asymptotic minima (when $\alpha \to \pi/2$, Fig. \ref{2Algo} bottom-left). The choice of $\lambda$ is hence not clear but for every reasonable value of $\lambda$, the criteria is optimized for $\alpha \to \pi/2$. Conversely $\lambda = 0.56$ is stable since its is $ 0.56 \simeq \lambda_{\alpha} \quad \forall \alpha \in [\pi/5 , \pi/2]$ moreover this is the optimal value we found for AT which is comforting. $\Xi_{MRA} < 0.01 =  \Xi_{AT}.10^{-1}$ (Fig. \ref{2Algo} bottom-right) which is very satisfactory.
In fact $\alpha = \pi/2$ means that the best tuning of the algorithm is "angle free". Either the vertex under consideration is a junction or there is at least an angle smaller than $\pi/2$. Consequently the condition on $\alpha$ is relaxed from $S_{\alpha}$ to $S = S_{\geq \pi/2}$.

The global minimum is the same for the two algorithms: $0.21$ but the branching coefficient is much smaller for MRA. Branches in streets are anecdotal when using $MRA$. 
We will in practice use the MRA in its maximal version that does not depend on the angle. 
\section{Street length distribution}
\subsection{Empirical street length fitting}
\begin{figure}
\includegraphics[scale=0.6]{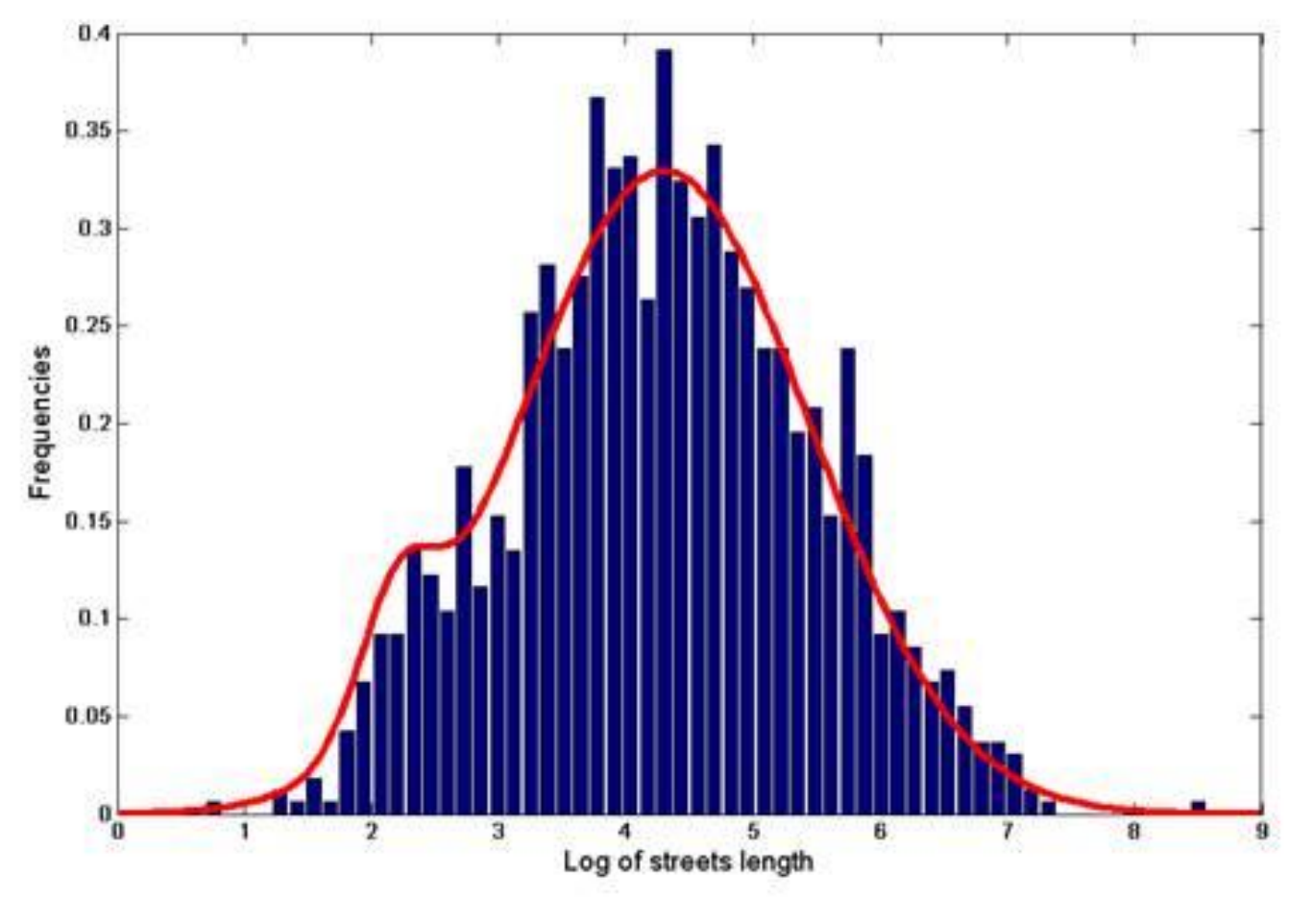}
\caption{\label{amiens} The distribution of the logarithm of street lengths in Amiens (France). The red curve it the fitting of this distribution by a mixture of two Gaussians.}
\end{figure}

\begin{figure}
\begin{center}
\includegraphics[scale = 0.13]{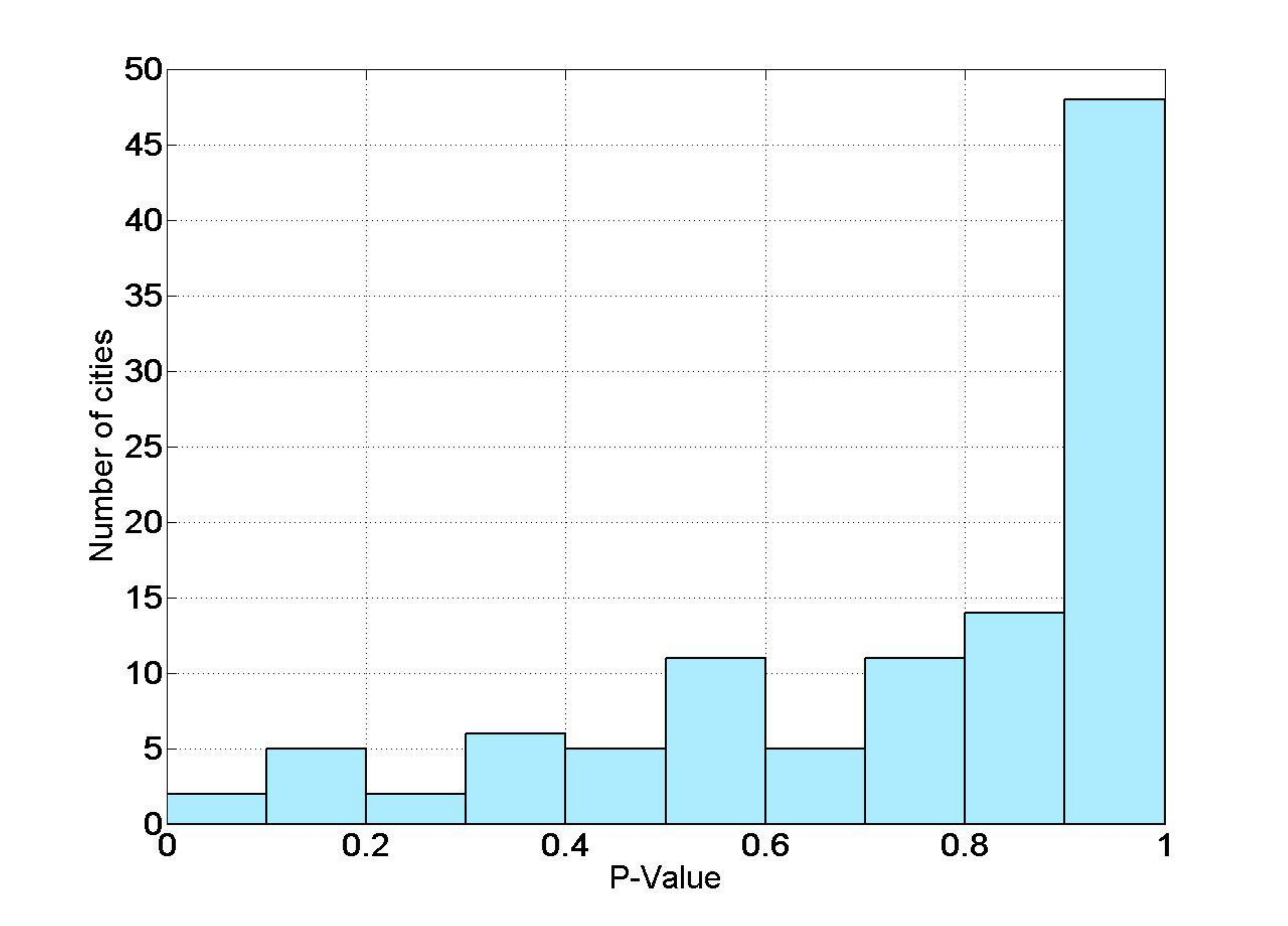}
\includegraphics[scale = 0.13]{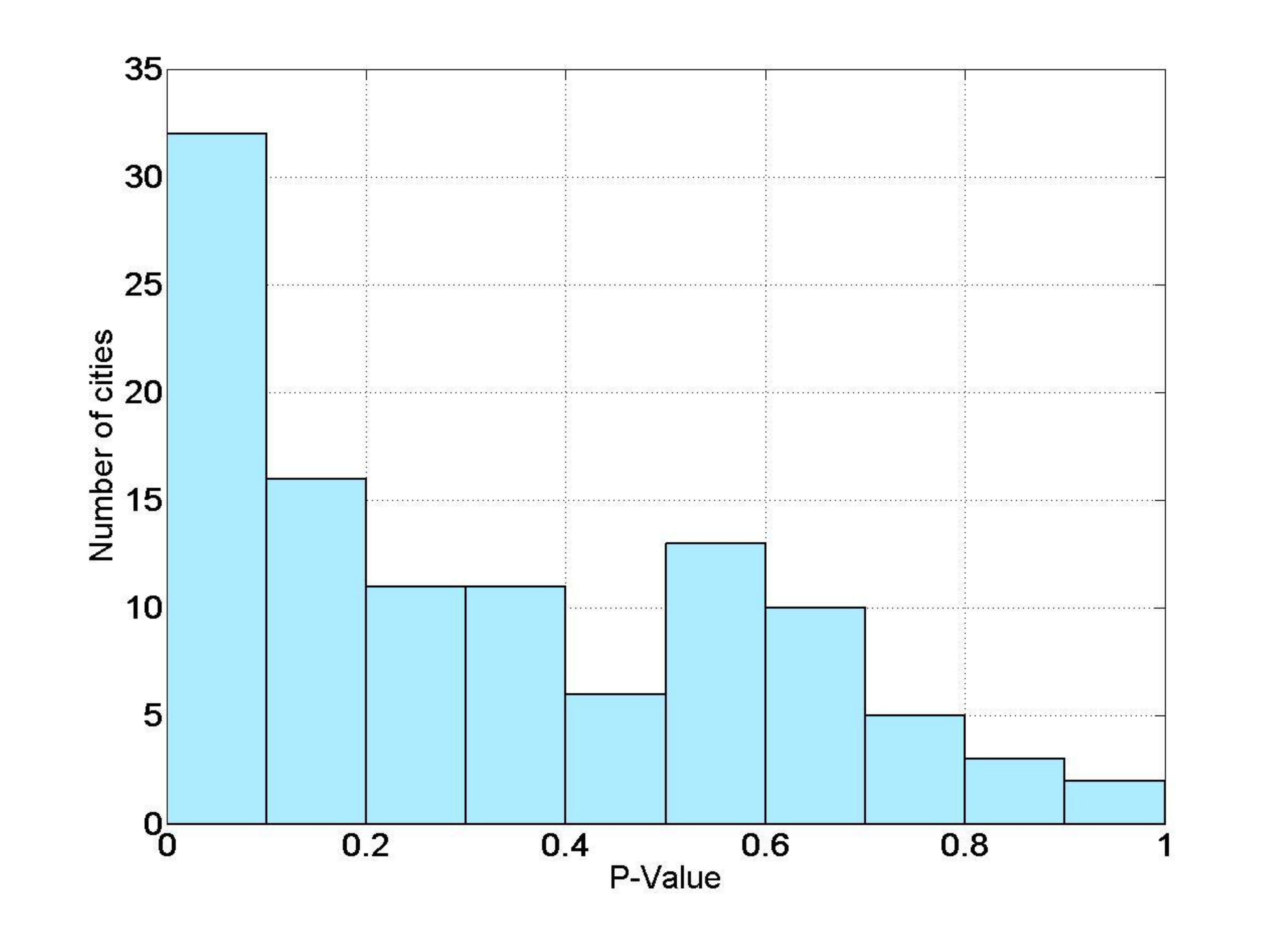}
\begin{tabular}{|c|c|c|c|c|c|}
\hline & Average & Min & Max & $>0.05$ &$>0.1$ \\ 
\hline Protocol 1 &$0.75$ & $0.042$ & $1$ & $99 \%$ & $98 \%$ \\ 
\hline Protocol 2 & $0.32$ & $1.9 \times 10^{-4}$ & $0.98$ &$77 \%$ &$70 \%$ \\ 
\hline 
\end{tabular} 
\caption{\label{tablePvalue} Main characteristics of the p-value distributions for the test "the distribution follows a mixture of log-normal" in 109 French Cities. Protocol 1 estimates parameters and performs a Kolomogorov -Smirnov test with the same data. Protocol 2 use the (randomly chosen) half of the streets to assess parameters and the other half to perform the test. }
\end{center}
\end{figure}

In a city, there are long streets assuring an efficient transportation system and small streets "fractal" distributed to provide habitation space. We thus expect that the distribution of street lengths $L$ exhibits a wide range of values or scales logarithmically. Fig. \ref{amiens} plots the distribution of the logarithm of street length in the French city Amiens. The global shape of this histogram suggests two maxima and two (different) normal tails. We assume that $\log L$ follows a mixture of two Gaussians (or similarly that $L$ follows a mixture of log-normal laws): 
\begin{equation}
\log{L} \sim p_-.\mathcal{N}(m_-, \sigma_-) + (1-p_-).\mathcal{N}(m_+, \sigma_+) 
\end{equation}
with $m_-<m_+$. The identification of this model has been performed with an Expectation Maximization algorithm.
Other cases show that the multi log normal distribution is robust even if it is possible to observe one or two maxima. For our whole data basis of French towns we calculate a bi-normal fitting of $L$ and calculated from a Kolomogorov - Smirnov test the p-value of this fitting: "\textit{L follows a mixture of two log normal laws}" against "\textit{L does not follow a mixture of two log normal laws}". We have chosen this test rather than a Chi-2 for its robustness to distribution supports. 
\\
In Protocol 1 we have for each city calculated the best parameters with an EM and calculated the P-value. It is often done this way in the literature. Nonetheless the statistics of the test is changed if parameters are estimated with the same data as for the test. In the normal case it remains the same asymptotically. We have not found a generalization to any distribution. 
\\
We propose a second protocol: since Kolomogorov -Smirnov is relevant from $100$ samples and our cities typically contain 500 to 1000 streets, we randomly divide each length distribution in two parts, used one to estimate parameters and the other to perform the test. The estimation and the test are done with less data and are less accurate. Results for both methods are summed-up in Fig. \ref{tablePvalue} and Tab. \ref{tablePvalue}.
The hypothesis is as relevant as the p-value is close to 1. Traditionally one considers that the hypothesis cannot be rejected if p-value$>0.1$. 
Let's focus on the second method. It is theoretically valid but needs randomization. From a realization to another the p-value of a particular city may highly change but the average p-value remains between $0.3$ and $0.4$. In $77 \%$ of cases the hypothesis is not rejected and in average the p-value is $0.32$ which is quite high. 
\subsection{Interpretation}
Log-normal laws are not rare in nature \cite{Limpert2001}. They appear in concentration of elements, latency periods of disease, rainfall, permeability in plant physiology... They are characteristic of multiplicative processes. We then could think that a city shapes by dividing in smaller blocks former blocks. This would lead to consider the city is the result of a division process as in \cite{Barthelemy2009}. \cite{Thale2009} recalls that for isotropic planar tessellations stable under iteration the length of the typical "I segment" (a street) is long-tailed but the result is not a log-normal. It is necessary to add a phenomenon to get the log-normal distribution. Maybe the extension of the city: people have a typical transportation length: $\lambda$. They accept to settle in a place where they have access to a constant volume of resources at a distance smaller than $\lambda$. Then when they cannot divide blocks they place at the exterior of the city into larger blocks. 
\\
To come to bimodality: this one does not appear on each city. A social science explanation is the following of several transportation mods along time or several populations build the city with two different policies (inhabitants and industries for instance).
\section{Conclusion}
We have presented a mathematical structure to consider a city not as a graph embedded in space but as a geometrical object. Similarly to Horton's method \cite{Horton1945} to break down tree structures in hydraulic, we have proposed a linear in time algorithm to recover streets in a general geometric graph. This algorithm might have depend on a parameter but reveals to be parameter free. Our algorithm is "more reliable" than the data. We define from city Hypergraph a new centrality: the simplest centrality \cite{Courtat2011a}. Contrary to other centralities such as betweeness, closeness or straightness that one varies softly and is side-effects free. It allows emphasizing important axes in a map and conversely to detect ill deserved zones. 
The behavior of street lengths leads to think of the city as the result of a morphogenetic process based on the duality extension / division of space \cite{Courtat2011}. 



\bibliographystyle{elsarticle-num}

\end{document}